\documentclass[prl,aps,twocolumn,superscriptadress,floatfix]{revtex4-1}

\usepackage{amssymb}
\usepackage{amsmath}
\usepackage{graphicx,bbm,color}
\allowdisplaybreaks
\usepackage{verbatim}
\usepackage{relsize}
\usepackage{hyperref}

\usepackage{dsfont}

\usepackage{dsfont}

\newcommand{\bgt}{\begin{itemize}}
\newcommand{\ent}{\end{itemize}}

\newcommand{\op}{\operatorname}

\newcommand{\lan}{\langle}
\newcommand{\ran}{\rangle}

\newcommand{\Tr}{\operatorname{Tr}}

\newcommand{\E}{\op{\mathbb{E}}}

\newcommand{\f}{\frac}

\newcommand{\bbm}{\begin{bmatrix}}
\newcommand{\ebm}{\end{bmatrix}}
\newcommand{\bes}{\begin{equation*}}
\newcommand{\ees}{\end{equation*}}
\newcommand{\be}{\begin{equation}}
\newcommand{\ee}{\end{equation}}
\newcommand{\beqy}{\begin{eqnarray}}
\newcommand{\eeqy}{\end{eqnarray}}
\newcommand{\beq}{\begin{eqnarray*}}
\newcommand{\eeq}{\end{eqnarray*}}

\newcommand{\bpm}{\begin{pmatrix}}
\newcommand{\epm}{\end{pmatrix}}

\long\def\symbolfootnote[#1]#2{\begingroup
\def\thefootnote{\fnsymbol{footnote}}\footnote[#1]{#2}\endgroup}

\begin{document}

\setkeys{Gin}{width=0.9\textwidth}

\title{Typical equilibrium state of an embedded quantum system}

\author{Gr\'egoire Ithier}      \affiliation{Department of Physics, Royal Holloway, University of London, TW20 0EX, Egham, United Kingdom}
\author{Saeed Ascroft}      \affiliation{Department of Physics, Royal Holloway, University of London, TW20 0EX, Egham, United Kingdom}
\author{Florent Benaych-Georges}     \affiliation{MAP5, UMR CNRS 8145 -- Universit\'e Paris Descartes, 75006 Paris France}

\begin{abstract}
We consider an arbitrary quantum system coupled non perturbatively to a large arbitrary and fully quantum environment.
 In [G. Ithier and F. Benaych-Georges, Phys. Rev. A \textbf{96}, 012108 (2017)] the typicality of the dynamics of such an embedded quantum system was established for several classes 
 of random interactions.
In other words, the time evolution of its quantum state does not depend on the microscopic details of the interaction.
  Focusing at the long time regime, we use this property to calculate analytically a new partition function characterizing the stationary state 
  and involving the overlaps between eigenvectors of a bare and a dressed Hamiltonian. This partition function provides a new thermodynamical ensemble which includes the 
  microcanonical and canonical ensembles as particular cases.
 We check our predictions with numerical simulations.
\end{abstract}

\maketitle






In what state of equilibrium can a quantum system be? Does this state have universal properties 
and what are the conditions for its emergence?
These questions are not new, dating even from the very birth of quantum theory\cite{vonNeumann} and
 are surprisingly open\cite{polkovnikov_colloquium_2011,eisert_quantum_2015}.
Indeed, the foundations of statistical physics still rely today on a static Bayesian point of view assuming the equiprobability of the accessible states defining the microcanonical ensemble. Assuming temperature and chemical potential can be defined then the canonical and grand canonical ensembles can be derived,
allowing to calculate all relevant macroscopic quantities
 in the thermodynamical limit~\cite{Kubo,LaudauLifschitz,Gemmer}.
In order to link theoretical predictions calculated  
with averages over these ensembles 
 to experimental quantities measured on a \textit{single} system, an assumption of ergodicity is made.
 Despite being broadly accepted, this assumption is not justified in a satisfactory manner
  (see, e.g., the discussion in Ref.~\cite{gemmer_distribution_2003}).  
Triggered by recent progress in the quantum engineering of mesoscopic systems
~\cite{bloch_quantum_2012,devoret_superconducting_2013},
 some theoretical progress has been achieved for attempting to explain thermodynamical equilibrium with a purely quantum point of view. 

From the early work of von Neumann on quantum ergodicity~\cite{vonNeumann,goldstein_normal_2010},
most theoretical studies aiming at understanding thermalisation as a quantum and universal~\cite{UnivNote} process
have focused on looking for 
  signatures of thermalisation on physical observables of large quantum systems~\cite{deutsch_quantum_1991,ReimannETH,reimann_typicality_2007,reimann_typical_2016}, 
  for instance with the Eigenstate Thermalisation Hypothesis (ETH) surmise~\cite{srednicki_chaos_1994,rigol_thermalization_2008,dalessio_quantum_2016}.
Instead of observables, one can also focus on the state of a system embedded in a larger one for which a ``canonical typicality'' property
has been established: the overwhelming majority of pure quantum states of the composite 
system are \textit{locally}\cite{LocalNote} canonical
\cite{goldstein_canonical_2006, tasaki_quantum_1998,popescu_entanglement_2006}.
This static ``typicality'' has been extended to the \textit{dynamics} of embedded quantum systems (two-level~\cite{Pastur2007}, 
four-level~\cite{bratus_qubits_2017} and arbitrary~\cite{ithier_dynamical_2017} quantum systems). We apply here this ``dynamical typicality'' 
property in order to calculate analytically and with full generality the stationary state of an embedded quantum system at long time. We find 
a new thermodynamical ensemble of purely quantum origin characterizing this state. This ensemble captures the microcanonical and the canonical ensembles as particular cases, and as such provides a quantum explanation for the Gibbs distribution.


We consider an arbitrary quantum system coupled to a large arbitrary quantum environment through a random interaction.
 We emphasize the fact that the initial state of this composite system can be chosen arbitrarily, in particular the environment \textit{does not} have to be in thermal equilibrium initially 
  nor the full composite system in the microcanonical situation.
  Dynamical typicality~\cite{Pastur2007,ithier_dynamical_2017} states that for almost all interaction Hamiltonian\cite{AlmostNote} the reduced density matrix of the system has a self-averaging property in the large environment limit\cite{PasturNote},
in other words, it follows a universal dynamics.
Despite this does not imply \textit{a priori} equilibration, since it can be consistent with sustained oscillations and revivals\cite{CommentRevivals}, this property has 
a very practical consequence. It allows to perform non perturbative analytical calculations with full generality, i.e. for \textit{arbitrary system, environment, and global initial state}, by justifying rigorously an averaging procedure over some randomness introduced \textit{only} at the level of the interaction Hamiltonian.
We apply this calculation framework here to study the state of the system at long but finite times, i.e. smaller than any recurrence time. 
Postponing all questions regarding the out of equilibrium dynamics to a further publication~\cite{PLnote}, we show that: \textit{if} the system converges towards a stationary state, then
this state is characterized by a new quantum partition function which can be calculated. This partition function
 relies on an average transition probability between states involving some purely quantum quantities: the fourth order moments of the overlap coefficients
  between eigenvectors of a bare and a dressed Hamiltonian.
We calculate this transition probability for several classes of random interactions.
 Then we calculate the probabilities of occupation of the states of the system a find a new thermodynamical ensemble more general than the microcanonical one.
 

\emph{Model Setup.}---
The setup is identical to\cite{ithier_dynamical_2017}: we consider a system $S$ in contact with an environment $E$, writing $\mathcal{H}_s$, $\mathcal{H}_e$ for their respective Hilbert spaces.
The total system $S+E$ is closed and its Hilbert space is the tensor product $\mathcal{H}=\mathcal{H}_s \otimes \mathcal{H}_e$ (with 
dimension $N =\dim \mathcal{H}_e . \dim \mathcal{H}_s$). The total or \textit{dressed} Hamiltonian $\hat{H}$ is the sum $\hat{H}=\hat{H}_s+\hat{H}_e+\hat{W}$ where $\hat{W}$ is an interaction term.
Eigenvectors of the ``bare'' Hamiltonian $\hat{H}_s+\hat{H}_e$ are written as $|\phi_n\rangle$ and are tensor products of eigenvectors $|\epsilon_s\rangle$ of $\hat{H}_s$ and
eigenvectors $| \epsilon_e\rangle$ of $\hat{H}_e$, with the eigenenergy $\epsilon_n=\epsilon_s+\epsilon_e$.
We write $|\psi_i\rangle$ for the dressed eigenvectors 
and $\{ \lambda_i \}_i$ the set of associated dressed eigenvalues. The state of $S+E$ is described by a density matrix $\varrho(t)$ which follows the well known relation 
$$ \varrho(t) =\hat{U}_t  \varrho(0) \hat{U}_t^\dagger \quad \text{with} \quad \hat{U}_t=e^{- \f{i}{\hbar} \hat{H} t}.$$
The state of the subsystem $S$ is 
described by a reduced density matrix:
$\varrho_s(t) = \Tr_e  \varrho(t),$ $\Tr_e$ being the partial trace with respect to the environment.
Decomposing the initial state $\varrho(0)$ on the bare eigenbasis $\{ | \phi_1 \ran,.., | \phi_N \ran \}$ and using linearity, we consider the matrix elements
 $\lan \phi_n | \hat{U}_t |\phi_m \rangle \langle \phi_p | \hat{U}^\dagger_t |\phi_q\ran$ 
in order to calculate $\varrho_s(t)$. By
expanding 
the evolution operator $\hat{U}_t$ over the dressed eigenbasis $\{|\psi_1\ran,...,|\psi_N\ran\}$:
$\hat{U}_t=\sum_{i} e^{-\frac{i}{\hbar } \lambda_it} | \psi_i \rangle \langle \psi_i|,$
these matrix elements 
 can be re-written as the $2$ dimensional Fourier transform of a product of four ``overlaps'' $\lan \phi_n | \psi_i \ran$:
\begin{eqnarray}
\label{FourierOverlaps}
  \nonumber 
  \lan \phi_n | \hat{U}_t    |\phi_m \rangle  \langle \phi_p |  \hat{U}_t^\dagger |\phi_q \ran =   \qquad \qquad \qquad  \qquad \qquad\\
   \sum_{i,j}
 e^{-\f{i}{\hbar}(\lambda_i-\lambda_j) t}
  \lan \phi_n  | \psi_i \ran \lan \psi_i | \phi_m\ran 
 \lan \phi_p | \psi_j  \ran  \lan  \psi_j | \phi_q\ran.
\end{eqnarray}
To calculate the expression in Eq.~\eqref{FourierOverlaps}, one needs an analytical formula for
the overlap coefficients
$\lan \psi_i | \phi_n \ran$ and the dressed eigenvalues $\lambda_i$, which are quantities
usually accessible in a perturbative framework only. In this Letter, we use a statistical method
for calculating these quantities in a \textit{non} perturbative setting and 
for arbitrary system and environment.

\emph{Introducing randomness.}---
The method relies on the hypothesis assumed for the interaction Hamiltonian:
we introduce \textit{deliberately} some randomness \emph{in and only in} the interaction $\hat{W}$
in order to perform calculations,
 knowing that this randomness actually will not matter in the large
 dimensionality limit ($\dim \mathcal{H}_e \to \infty$) due to the typicality of the dynamics
\cite{ithier_dynamical_2017}.
This randomness should be compatible with some macroscopic constraints: $\hat{W}$ ``centered'' i.e. $\Tr(\hat{W})=0$ and with fixed spectrum variance 
$\sigma_w^2=\Tr(\hat{W}. \hat{W}^\dagger)/N$ independent of $N$.
 Then regarding the symmetry class of the randomness, we will assume $\hat{W}$ to be either a Wigner band random matrix (WBRM)\cite{WBRMNote}
or a randomly rotated matrix
 (RRM i.e. of the type $ \hat{U}.\hat{Q}.\hat{U}^\dagger $ with $\hat{Q}$ real diagonal fixed and $\hat{U}$ unitary or orthogonal Haar distributed). 
The WBRM ensembles are convenient for modeling interactions in heavy atoms and nuclei\cite{FlaumbaumStructureCompoundState1994,fyodorov_wigner_1996,borgonovi_quantum_2016}. 
The sparsity of WBRM 
comes from the finite energy range of the interaction.
On the other hand, RRM ensembles are dense, which contradicts the \textit{a priori} two body nature
 of the interaction, but provides a convenient way for modeling the local spectral statistics of more physical interaction Hamiltonians\cite{PorterBook,brody_random-matrix_1981,Mehta,ericson_statistical_1960,borgonovi_quantum_2016}.

Then we focus on the reduced density matrix: $\varrho_s(t)=\Tr_e \varrho(t)$ and consider it as a function of the interaction $\hat{W}$,
 keeping all other parameters constant (time, spectra of $S$ and $E$, initial state).
From \cite{ithier_dynamical_2017}, we know that this function undergoes
 for the random matrix ensembles considered, 
 a phenomenon known as the ``concentration of measure"\cite{MR1387624}: as $\dim \mathcal{H}_e \to \infty$,
$\varrho_s(t)$ is getting very close to its mean value which provides the typical dynamics.
Consequently, we can compute an approximate $\varrho_s(t)$ simply by averaging:
$\varrho_s(t)= \Tr_e( \varrho(t) ) \approx \mathbb{E} [\Tr_e (\varrho(t))] = \Tr_e \left( \E[\varrho(t)]\right)$, where $\mathbb{E}$ is the average over the set of interaction Hamiltonians considered.
We are thus led to consider the average of Eq.~\eqref{FourierOverlaps}.

\emph{Statistics of the overlaps.}--- We will now focus specifically on the stationary regime at long times. 
Under the hypothesis assumed on the statistics of the interaction  (WBRM and RRM ensembles) 
the dressed eigenvalues $\{\lambda_1,...,\lambda_N\}$ undergo level repulsion and, as such, are non degenerate. This implies that the time independent terms are provided by the case $i=j$ in the summation in Eq.\eqref{FourierOverlaps} averaged over $\hat{W}$:
\begin{equation}
\label{LongTimeLimit}
 \sum_{i} \E[  \lan \phi_n  | \psi_i \ran \lan \psi_i | \phi_m\ran 
 \lan \phi_p | \psi_i  \ran  \lan  \psi_i | \phi_q\ran].
\end{equation}

The time dependent regime (given by the summation over $i$ and $j$ such that $ i \neq j$) 
is outside the scope of this article.
We will assume this regime to be damped (see \cite{genway_dynamics_2013} for $\hat{W}$ in the WRBM ensemble),
 without revivals\cite{CommentRevivals} at least on
the largest time scale of this model ($1/D$ where $D$ is the mean level spacing of the
dressed Hamiltonian) such that considering a stationary regime is meaningful over this time scale.

We first single out the non zero cases for the fourth order moments of the overlap
 coefficients: $\E[  \lan \phi_n  | \psi_i \ran \lan \psi_i | \phi_m\ran 
 \lan \phi_p | \psi_i  \ran  \lan  \psi_i | \phi_q\ran]$ which are
when ($n=m$ and $p=q$) or  when ($n=q \text{ and } m=p$)~\cite{SuppMatNote}.
 The former case is involved in the asymptotic value of the off-diagonal terms of $\varrho_s(t)$ i.e. the quantum coherences of the state of $S$, which can be shown to be zero as expected in the limit $t \to \infty$~\cite{SuppMatNote}.
In the following, we focus on the later case ($n=q \text{ and } m=p$) which governs the dynamics of the diagonal terms of $\varrho(t)$ and $\varrho_s(t)$, i.e. the probabilities of occupation.

\emph{Average transition probability}.---
 We define from Eq.\eqref{LongTimeLimit}  with $n=q$ and $m=p$, an \textit{average} transition probability $\bar{p}_{m \to n}$ 
from an initial state $|\phi_m \ran$ at $t=0$ to a final state $|\phi_n \ran$ at $t \to \infty$:
\begin{eqnarray}
\bar{p}_{m \to n} = \sum_i \E[ |\lan \phi_n | \psi_i \ran |^2  |\lan \phi_m | \psi_i\ran |^2 ].
  \label{SumI1} 
\end{eqnarray}
Such sum provides quantitatively how $|\phi_n \ran$ is accessible from $|\phi_m \ran$
and has been considered, e.g., numerically in the context of random two body
interactions (TBRI) ensembles~\cite{flambaum_correlations_1996} and analytically for some specific systems:
quantum walkers\cite{JMLuckQWalkers,luck2016investigation,JMLuck2017}.
 The particular case $m=n$ provides the return probability whose reciprocal $1/\bar{p}_{n \to n}$ is the so-called purity~\cite{linden_quantum_2009,linden_speed_2010,short_equilibration_2011,ikeda_how_2015}. 
The leading order of $\bar{p}_{m \to n}$ is given by
$\sum_i \E[ |\lan \phi_n | \psi_i \ran |^2 ] \E[ |\lan \phi_m | \psi_i \ran |^2 ]$
and involves the second order moment of the overlaps $\E[|\lan \phi_n | \psi_i \ran |^2]$. 
This quantity, multiplied by $\rho$, is called the \textit{local} density of states (LDOS)
 and quantifies how much a bare
  eigenvector is delocalized or hybridized with the dressed eigenbasis
and  has already been considered in
various contexts (nuclear physics\cite{BreitWigner1936,wigner_characteristic_1955,wigner_characteristics_1957},
molecular physics\cite{rice_predissociation_1933}, atomic physics\cite{FlaumbaumStructureCompoundState1994}, thermalisation\cite{deutsch_quantum_1991}, quantum chaos\cite{FlaumbaumStructureCompoundState1994}, financial data analysis\cite{allez_eigenvectors_2013,allez_eigenvectors_2014}, see also the review in \cite{borgonovi_quantum_2016}) 
 for various cases of  $\hat{H}_0$ and $\hat{W}.$
  It has the following
typical  shape:
 \begin{eqnarray}
\label{SecondMoments}
 \E[ |\lan \phi_n | \psi_i \ran |^2 ] \approx
\f{ f\left(\bar{\lambda}_i - \epsilon_n \right) }{   \int \rho_{s+e}(\epsilon) f( \bar{\lambda}_i - \epsilon) d\epsilon}
\end{eqnarray}
where  $\rho_{s+e}$ is the bare density of states, $\bar{\lambda}_i$ is the mean of the dressed eigenvalue $\lambda_i$
and $f$ is a function peaked around zero with a typical width $\Gamma$. The denominator is here for the purpose of normalisation.
For most models of $\hat{H}_0$ and $\hat{W}$,
the function $f$ is a Lorentzian reminiscent of the Breit-Wigner law with a generalized Fermi Golden
rule rate $\Gamma=\pi \sigma_w^2 \rho/N$, $\rho$ being the dressed density of states (see, e.g., Ref.~\cite{borgonovi_quantum_2016}). 
Interestingly, such a Lorentzian shape has been shown to preclude thermalisation in closed quantum systems made of interacting particles
as far as observables of these systems are concerned\cite{santos_chaos_2012,torres-herrera_general_2014,borgonovi_quantum_2016}. 
However, regarding the problem we are interested in: a quantum system coupled to a large environment, it is important to stress that this Lorentzian shape
 does \textit{not} preclude thermalisation, as we observe numerically (see Fig.1) and as far as the state of this embedded system is concerned. 
 This point is rather subtle and its explanation involves dynamical typicality (see \cite{SuppMatNote} for a detailed discussion).
Finally, we emphasize that the subsequent calculation can also be performed using other shapes (see \cite{SuppMatNote} for details and a short review of possible LDOS). In
 particular, our derivation can be applied to a Gaussian LDOS, relevant if $\hat{W}$ if enforces a two body nature of the interaction (TBRI) 
  \cite{flambaum_correlations_1996,Kota}.

 Assuming the interaction to be non perturbative, i.e. the mean level spacing $D$ is much smaller than the width  $\Gamma$ and consequently the bare eigenvector $|\phi_m \ran$ is
  delocalized over several ($\approx \Gamma \rho$) dressed eigenvectors, then one can proceed further with the calculation of $\bar{p}_{m \to n}$ by using
   a continuous approximation for the summation ($\sum_i \leftrightarrow \int \rho(\lambda) d \lambda $).
   The transition probability is then given by:
\begin{eqnarray}
\bar{p}_{m \to n}   \approx \f{g\left( \epsilon_m - \epsilon_n \right)}{ \int   \rho_{s+e}(\epsilon) g(\epsilon_m-\epsilon) d\epsilon }
  \label{SumI2} 
\end{eqnarray}
where $g= f \ast f$ is the convolution of $f$ with itself 
and with a typical width $\Gamma'$. For instance, if the LDOS is Lorentzian (resp. gaussian)
  then $g$ is also a Lorentzian (resp. gaussian) with a width $\Gamma' = 2 \Gamma$ (resp. $\Gamma' =\sqrt{2} \Gamma$).
At this stage, one should note that Eq.\eqref{SumI2} is in sharp contrast with the microcanonical hypothesis of equiprobability of the accessible states. 
We have performed numerical simulations for $\bar{p}_{m \to n}$ with $\hat{W}$ in the Gaussian orthogonal ensemble (GOE) and found a satisfactory agreement with
 our prediction\cite{SuppMatNote}.

\emph{Typical asymptotic state.}---
Finally, to perform the partial trace and get $\varrho_s(t)$,
we recall the final state $|\phi_n \rangle= |\epsilon_{s} \rangle |\epsilon_{e} \rangle$
and sum Eq.\eqref{SumI2} over $\epsilon_{e}$ using a continuous approximation: 
$\Tr_e =\sum_{\epsilon_{e}} \leftrightarrow \int d\epsilon  \; \rho_e(\epsilon)$.  
This provides the main result of this paper: for an initial state $\varrho(0)=|\phi_m \ran \lan \phi_m |$,
the long time stationary state of $S$ is distributed according to
\begin{eqnarray}
\label{ProbaS2}
p_{\epsilon_{s}} =  \lim_{t \to \infty} \langle \epsilon_{s} | \varrho_s(t) | \epsilon_{s} \rangle \approx 
\dfrac{ \int  \rho_e(\epsilon_e) g\left( \epsilon_m - \epsilon_s-\epsilon_e\right) d\epsilon_e}{
\int   \rho_{e+s}(\epsilon) g(\epsilon_m-\epsilon) d\epsilon}. \qquad
 \end{eqnarray}
 The denominator is the convolution of the \textit{bare} density of states by the transition probability $g$ which provides the effective number of bare states accessible from the initial  $|\phi_m\ran $. Such a quantity $\int   \rho_{e+s}(\epsilon) g(\epsilon_m-\epsilon) d\epsilon$ enforces the normalization condition and can be considered as a new partition function.
The numerator is the convolution of the \textit{environment} density of states by the transition probability $g$ and provides the effective number of accessible states
   such that $S$ is in the state of energy $\epsilon_s$. 
   The probability of occupancy is the ratio of these two numbers.
Let us now consider the case of intermediate coupling.

 \emph{Intermediate coupling.}---
 A temperature can be defined by $\beta=\frac{1}{k T}= \frac{d \ln  \rho_e}{d \epsilon} $. 
Assuming a good decoupling between the micro ($D=1/\rho$), meso ($\Gamma'$) and macro ($kT$) energy scales: $D \ll \Gamma' \ll kT$, and considering all energies $\epsilon_n, \epsilon_m$ to 
 be inside the bulk of the spectrum, then 
 the function $g$ in Eq.\eqref{ProbaS2} can be approximated by a Dirac function which is "sampling" $\rho_e(\epsilon)$ at $\epsilon_{e}=\epsilon_m-\epsilon_{s}$ and simplifying Eq.\eqref{ProbaS2} for
\begin{equation}
\label{ProbaS3}
p_{\epsilon_{s}} \approx  \frac{\rho_e (\epsilon_m-\epsilon_{s})}{ \rho_{s+e}(\epsilon_m)}.
\end{equation}
We are recovering here the same prediction as the one resulting from a microcanonical ensemble defined \textit{locally} in energy, i.e. by assuming the equiprobability of all bare eigenstates inside a small energy window centered around the initial energy $\epsilon_m$. This prediction is checked numerically on Fig. 1.
 It is important to stress that we recovered this prediction with a purely
quantum point of view: 
from the geometrical relation between the eigenvectors of the bare and dressed
Hamiltonians. 
Note that by assuming the environment to be macroscopic, i.e. $kT$ does not depend on energy on a wide range and consequently $\rho_e$ scales exponentially with energy, one can recover the canonical ensemble prediction following the usual derivation\cite{Kubo}:
 $$p_{\epsilon_s}\approx  \f{\rho_e(\epsilon_m-\epsilon_{s})}{\rho_{s+e}(\epsilon_m)}
= \f{\rho_e(\epsilon_m) e^{-\beta \epsilon_{s}}}{\sum_{\epsilon_{s'} }\rho_e(\epsilon_m) e^{-
\beta \epsilon_{s'}}} \approx \f{e^{-\beta \epsilon_{s}}}{Z_\beta} $$
with $ Z_\beta=\sum_{\epsilon_s} e^{-\beta \epsilon_s}  $ the canonical partition function. 
In other words, the Boltzmann distribution is a particular case of the more general distribution provided by Eq.\eqref{ProbaS2} whose origin is quantum.

 \begin{figure}    
\includegraphics[width=0.46\textwidth]{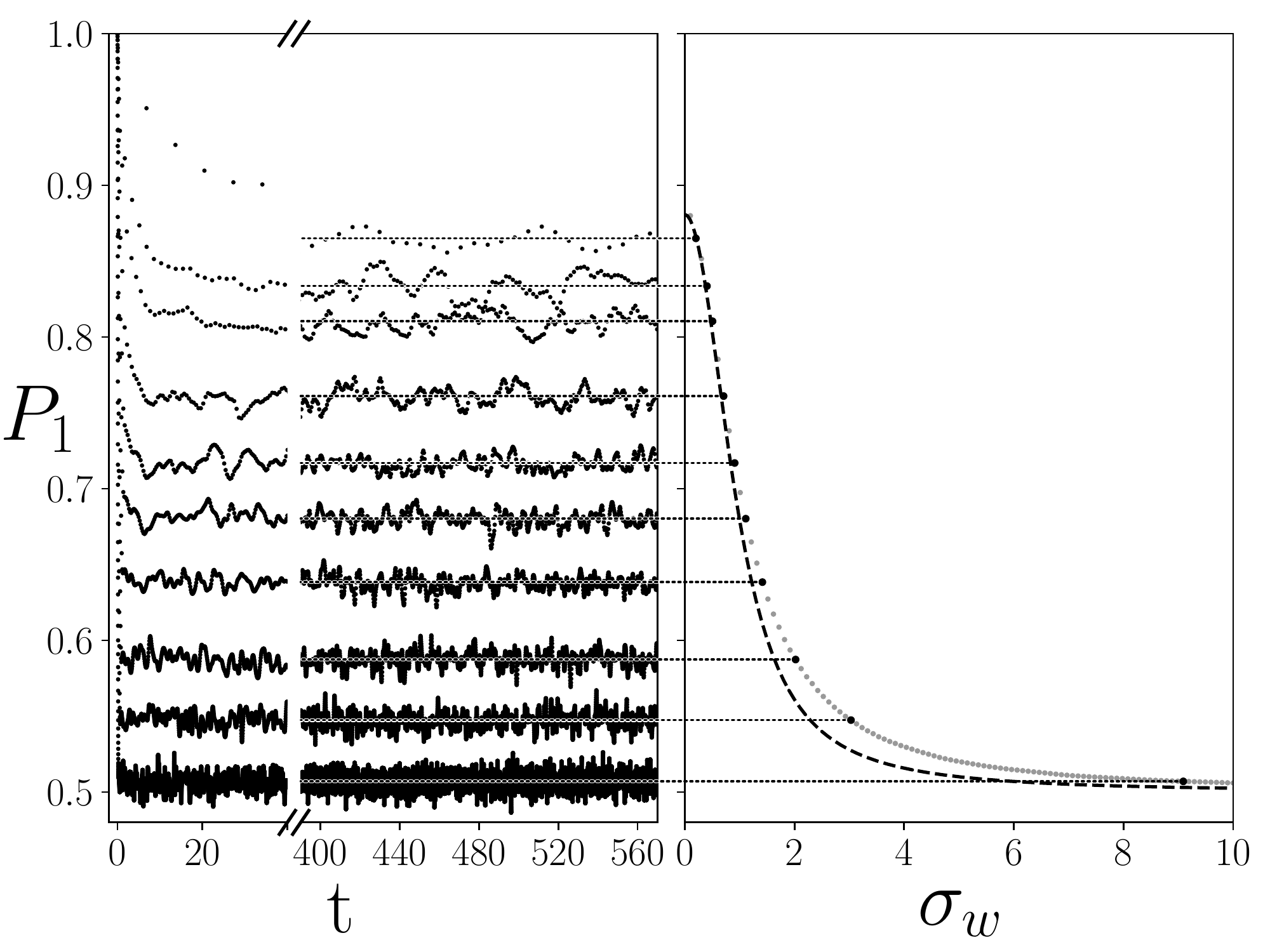}   
\caption{  \textbf{ Crossover from a \textit{local} microcanonical ensemble to a \textit{global} microcanonical ensemble.}
  We consider here numerically the particular case of a two level system $S$ (gap $\Delta=2$) coupled to an environment having a Gaussian density of states 
  (standard deviation $\sigma_e=1$) through an interaction $\hat{W}$ in the GOE ensemble.
  We plot $P_1$ the probability for the system to be in its excited state as a function of time (left panel) and then the long time average of $P_1$ as a function of interaction strength $\sigma_w=\sqrt{\Tr(W^2)/N}$ (right panel).
The environment Hilbert space dimension is set to $\dim \mathcal{H}_e=4096$ (so that the total Hilbert space dimension is $N=8192$), the initial state to $|1_s \ran \lan 1_s | \otimes |2048_e \ran \lan 2048_e |$ (i.e. middle of the spectrum for $E$: $\epsilon_e \approx 0$) and we numerically integrate the Schr\"{o}dinger equation 
for different values of $\sigma_w$.
After a transient regime at short times ($t \lesssim 40$), a stationary regime takes place.
As the interaction strength increases, the time average value of $P_1$ goes from a \textit{local} microcanonical prediction $ P_1 \approx  \rho_e(\Delta)/(\rho_e(\Delta) + \rho_e(0))\approx 0.87$ to a \textit{global} microcanonical prediction: $P_1 \approx 0.5$. 
The analytical prediction for this crossover (in dash) is given by Eq.\eqref{ProbaS2} which can be calculated analytically in this case: it is the convolution of a gaussian DOS with a Lorentzian transition probability $g$ giving the Voigt function\cite{Voigt} (see \cite{SuppMatNote} for details).
Note that in the intermediate coupling regime, the LDOS (and consequently the average transition probability $\bar{p}_{m \to n}$) is of the Breit-Wigner type and does \textit{not} preclude thermalization (see discussion in \cite{SuppMatNote} for details).
}
\label{Fig2}
\end{figure}

\emph{Strong coupling.}---
If the coupling is strong enough that $\Gamma'  \gtrsim kT$
then the transition probability $g$ cannot be approximated by a Dirac function and its finite width must be taken into
account in the convolution in Eq.\eqref{ProbaS2}. From this convolution effect, one should expect a \textit{decrease} of contrast in the probability distribution of $S$ when the
 interaction strength is increased: the equilibrium probability then undergoes a continuous crossover from the \textit{local} microcanonical ensemble prediction we described earlier (i.e. equiprobability over a small energy shell of accessible states around initial energy) to a \textit{global} microcanonical ensemble prediction (i.e. all bare state are accessible and equiprobable).
The convolution in Eq.\eqref{ProbaS2} can be done analytically e.g. when $\rho_{e}$ is Gaussian and $g$ is Lorentzian: one obtains the Voigt distribution, relevant in atomic spectrocopy when a natural linewidth is broadened by the Doppler effect\cite{Voigt}. 
We check numerically these predictions on Fig.~\ref{Fig2} and find a satisfactory agreement.

 Finally, we stress that the above results are valid for an initial state $ | \phi_m \ran \otimes \lan \phi_m | = |\epsilon_s \ran \lan \epsilon_s | \otimes |\epsilon_e \ran \lan \epsilon_e| $ and can be extended by linearity to any initial state, \textit{pure or not}: the extra diagonal terms (i.e. of the type $|\phi_m \ran \lan \phi_p |$ with $m \neq p$) do not contribute\cite{SuppMatNote}, only the diagonal ones contribute.
Therefore the stationary state of S  is the weighted average of
  Eq.~\eqref{ProbaS2} by the initial energy distribution of the composite system.

\emph{Conclusion.}--- 
We showed that the stationary properties of an embedded quantum system
  are encoded in the geometric relation between the eigenvectors of 
  a bare and a dressed Hamiltonian, more precisely in the fourth order moments of the
   overlaps between their eigenvectors. This fact provides a purely quantum way to define a
    new partition function which can be calculated thanks to
      dynamical typicality\cite{ithier_dynamical_2017}.   
In the intermediate coupling case $D \ll \Gamma' \ll kT$, this partition function simplifies to
the prediction of a local microcanonical ensemble defined on a small energy window around the initial energy.
 In the strong coupling regime (i.e. $D \ll kT \lesssim \Gamma'$), one gets a more
general ensemble which depends on the interaction strength and leads to a loss of contrast of
the probabilities of occupation (i.e. a convergence towards global equiprobability).
We considered here two random matrix ensembles for the interaction which have broad applicability. Our framework could be used with other interaction Hamiltonian ensembles
 (e.g. conserving some set of observables or enforcing the two body nature of the interaction) as soon as dynamical typicality is shown to be verified and a local density of states is available.

\emph{Acknowledgements.}---
We wish to thank D. Esteve and H. Grabert for their critical reading of the manuscript, their support and the numerous discussions, as well as J.-M. Luck, B. Cowan and X. Montiel for their useful comments and the discussions.


\begin{thebibliography}{60}
\bibitem [{Uni()}]{UnivNote}%
  \BibitemOpen
  \href@noop {} {}\bibinfo {note} {By universal, we mean independent of the
  specific Hamiltonian considered.}\BibitemShut {Stop}%
\bibitem [{Loc()}]{LocalNote}%
  \BibitemOpen
  \href@noop {} {}\bibinfo {note} {By locally we mean after taking a partial
  trace to get the state of the subsystem.}\BibitemShut {Stop}%
\bibitem [{Alm()}]{AlmostNote}%
  \BibitemOpen
  \href@noop {} {}\bibinfo {note} {This ``almost all'' is relative to the
  probability measure defined on the random matrix ensemble
  considered.}\BibitemShut {Stop}%
\bibitem [{Com()}]{CommentRevivals}%
  \BibitemOpen
  \href@noop {} {}\bibinfo {note} {Because of the finite dimension of the
  Hilbert space, this regime might present quantum recurrences and
  revivals.}\BibitemShut {Stop}%
\bibitem [{WBR()}]{WBRMNote}%
  \BibitemOpen
  \href@noop {} {}\bibinfo {note} {WBRM are of the type $W_{i,j}= a((i-j)/b)
  Y_{i,j}$ where $Y$ is a Wigner Random Matrix and $a(x)$ is a deterministic
  band profile, ``b'' being the bandwidth.}\BibitemShut {Stop}%
\end{thebibliography}

%

\end{document}